\PassOptionsToPackage{pdfpagelabels=false}{hyperref}

\documentclass[fleqn,usenatbib]{mnras}

\usepackage{graphicx}
\usepackage{amsmath}
\usepackage{amssymb}
\usepackage[dvipsnames]{xcolor}  
\usepackage[figure,figure*,table,table*]{hypcap}
\usepackage[normalem]{ulem}

\input{macros.sty}

\title[Cosmological Evidence Modelling]
{Cosmological Evidence Modelling: a new simulation-based approach to constrain cosmology on non-linear scales}
\author[J.~U.~Lange et al.]
{Johannes~U.~Lange$^1$\thanks{email: johannesulf.lange@yale.edu}, Frank~C.~van~den~Bosch$^1$, Andrew~R.~Zentner$^2$,\newauthor Kuan~Wang$^2$, Andrew P. Hearin$^3$ and Hong Guo$^4$\\
	$^1$Department of Astronomy, Yale University, P.O. Box 208101, New Haven, CT 06511, USA\\
	$^2$Department of Physics and Astronomy \& Pittsburgh Particle Physics, Astrophysics, and Cosmology Center (PITT PACC),\\University of Pittsburgh, Pittsburgh, PA 15260, USA\\
	$^3$Argonne National Laboratory, Argonne, IL 60539, USA\\
	$^4$Key Laboratory for Research in Galaxies and Cosmology, Shanghai Astronomical Observatory, Shanghai 200030, China\\}

\pubyear{2019}

\begin{document}

	\date{Accepted xxx. Received xxx}
	
	\label{firstpage}
	\pagerange{\pageref{firstpage}--\pageref{lastpage}}
	
	\maketitle
	
	\begin{abstract}
	    Extracting accurate cosmological information from galaxy-galaxy and galaxy-matter correlation functions on non-linear scales ($\lta 10 \Mpch$) requires cosmological simulations. Additionally, one has to marginalise over several nuisance parameters of the galaxy-halo connection. However, the computational cost of such simulations prohibits naive implementations of stochastic posterior sampling methods like Markov chain Monte Carlo (MCMC) that would require of order $\mathcal{O}(10^6)$ samples in cosmological parameter space. Several groups have proposed surrogate models as a solution: a so-called emulator is trained to reproduce observables for a limited number of realisations in parameter space. Afterwards, this emulator is used as a surrogate model in an MCMC analysis. Here, we demonstrate a different method called Cosmological Evidence Modelling (CEM). First, for each simulation, we calculate the Bayesian evidence marginalised over the galaxy-halo connection by repeatedly populating the simulation with galaxies. We show that this Bayesian evidence is directly related to the posterior probability of cosmological parameters. Finally, we build a physically motivated model for how the evidence depends on cosmological parameters as sampled by the simulations. We demonstrate the feasibility of CEM by using simulations from the {\sc Aemulus} simulation suite and forecasting cosmological constraints from BOSS CMASS measurements of redshift-space distortions. Our analysis includes an exploration of how galaxy assembly bias affects cosmological inference. Overall, CEM has several potential advantages over the more common approach of emulating summary statistics, including the ability to easily marginalise over highly complex models of the galaxy-halo connection and greater accuracy, thereby reducing the number of simulations required.
	\end{abstract}
	
	
	\section{Introduction}
	
	The distribution of galaxies and matter encodes vital clues about cosmology. The advent of large galaxy surveys has propelled our ability to constrain cosmological parameters from observations of the late Universe. Examples include the \textit{Sloan Digital Sky Survey} \citep[SDSS,][]{Abazajian_09}, the \textit{Baryon Oscillation Spectroscopic Survey} \citep[BOSS,][]{Reid_16}, the Dark Energy Survey \citep[DES,][]{DES_18a} or the upcoming \textit{Dark Energy Spectroscopic Instrument} \citep[DESI,][]{DESI_16} survey.
	
	The study of the distribution on small, $\sim \mathrm{Mpc}$ scales potentially holds the strongest statistical constraining power due to its high signal-to-noise measurements but also comes with unique challenges. First, the matter density contrast $\delta$ on such small, so-called non-linear scales strongly deviates from zero, $|\delta| \gg 0$, prohibiting analytic predictions from perturbation theory that are possible in the linear and mildly non-linear regime. Secondly, the cross correlation coefficient of galaxy and matter $r_{\rmg \rmm}$ deviates from unity \citep[e.g.][]{Seljak_00, Cacciato_12, Wibking_19}. This necessitates the need to model several aspects of the relationship between galaxies and the dark matter haloes harbouring them. This relationship between galaxies and haloes is colloquially known as the galaxy-halo connection. Based on the above considerations, it is challenging to obtain accurate cosmological inferences from the study of small scales and many studies therefore resort to the linear and mildly non-linear regime \citep[e.g.][]{Mandelbaum_13, Beutler_17, DES_18a}.
	
	To model the matter distribution on small scales, many previous works \citep[see e.g.][]{Cacciato_09, vdBosch_13, Cacciato_13, More_15} employed empirically derived, analytic fitting functions \citep[see e.g.][]{Smith_03, Tinker_08, Tinker_10, Klypin_16} obtained from a small number of numerical simulations of structure formation. However, it is evident that the precision of these fitting formulas is insufficient given the accuracy of upcoming measurements \citep[see e.g.][]{vdBosch_13}. Thus, future works need to employ a fully simulation-based approach to use non-linear scales to constrain cosmology. At the same time, one also needs to allow for complex models of the galaxy-halo connection and propagate our uncertainties in those models into the cosmological analysis. As we will discuss below, a popular approach to obtain cosmological constraints in the nonlinear regime is to calibrate surrogate models for simulation-based predictions of observable summary statistics \citep[e.g.,][]{Heitmann_06, Habib_07, Heitmann_10}. The goal of this paper is to develop an alternative method based on Bayesian evidence that relaxes important limitations on the level of modelling complexity that can be achieved with conventional implementations of the surrogate modelling approach. We test this new method, which we call Cosmological Evidence Modelling (CEM), by forecasting constraints from redshift-space clustering of BOSS CMASS galaxies.
	
	Our paper is organised as follows. In section \ref{sec:methods}, we briefly review the statistical background, discuss surrogate modelling, and outline our new CEM method, the detailed elements of which are described in section~\ref{sec:ingredients}. Next, we perform an analysis of mock BOSS CMASS data in section \ref{sec:application} and specifically discuss the role of galaxy assembly bias in \ref{sec:gab}. Finally, we summarise our findings in section \ref{sec:conclusion}. Throughout this work, $\log$ refers to the decadic logarithm and $\ln$ to the natural logarithm.
	
	\section{Statistical methods}
	\label{sec:methods}
	
	Generally, our goal is to constrain a cosmological model, denoted by $\mathcal{C}$, given a certain set of observational data $\mathbf{D}$, i.e. a set of summary statistics extracted from galaxy populations. However, in large-scale galaxy surveys, the galaxy-halo connection $\mathcal{G}$ describing how galaxies occupy dark matter haloes is another latent variable that needs to be modelled. Thus, in order to obtain observational constraints on cosmology we need to marginalise over our uncertainty in $\mathcal{G}$. Using Bayes' theorem, the posterior probability of $\mathcal{C}$ and $\mathcal{G}$ given some observational data $\mathbf{D}$ is equal to
	\begin{equation}
	    P \left( \mathcal{C}, \mathcal{G} | \mathbf{D} \right) = \frac{P\left( \mathbf{D} | \hat{\mathbf{D}} \left( \mathcal{C}, \mathcal{G} \right) \right) P(\mathcal{C}) P(\mathcal{G})}{\mathcal{Z} (\mathbf{D})}.
	    \label{eq:bayes_full}
	\end{equation}
	In the above equation $\hat{\mathbf{D}} \left( \mathcal{C}, \mathcal{G} \right)$ is the expectation value for $\mathbf{D}$ given a certain model for cosmology and the galaxy-halo connection. $P( \mathbf{D} | \hat{\mathbf{D}} )$ is then the likelihood to obtain $\mathbf{D}$ given the expectation value $\hat{\mathbf{D}}$ and depends on the accuracy of the measurement of $\mathbf{D}$. $P(\mathcal{C})$ and $P(\mathcal{G})$ denote prior knowledge on cosmology and the galaxy-halo connection, respectively. Finally, $\mathcal{Z} (\mathbf{D})$ is the Bayesian evidence of the data. It can be obtained by integrating over the parameter space of $\mathcal{C}$ and $\mathcal{G}$,
	\begin{equation}
	    \mathcal{Z} (\mathbf{D}) = \int \int P\left( \mathbf{D} | \hat{\mathbf{D}} \left( \mathcal{C}, \mathcal{G} \right) \right) P(\mathcal{C}) P(\mathcal{G}) \rmd \mathcal{C} \, \rmd \mathcal{G},
	\end{equation}
	and ensures proper normalisation. There exist several numerical techniques, e.g. MCMC, to sample from the above distribution without the need to know its normalisation $\mathcal{Z} (\mathbf{D})$. This ultimately allows us to obtain posterior samples for $\mathcal{C}$ and $\mathcal{G}$. Finally, we can obtain observational constraints on cosmology by marginalising over $\mathcal{G}$,
	\begin{equation}
	    P\left( \mathcal{C} | \mathbf{D} \right) = \int P\left( \mathcal{C}, \mathcal{G} | \mathbf{D} \right) \rmd \mathcal{G}.
	\end{equation}
	The above quantity is the major result that we seek to obtain from large-scale structure galaxy surveys. Ultimately, to sample from the posterior probability of $\mathcal{C}$, we need to predict the expectation value $\hat{\mathbf{D}} \left( \mathcal{C}, \mathcal{G} \right)$. However, to stochastically sample from the posterior probability of both $\mathcal{C}$ and $\mathcal{G}$ in Eq.~\eqref{eq:bayes_full}, we need of order $\mathcal{O}(10^6)$ evaluations of $\hat{\mathbf{D}} \left( \mathcal{C}, \mathcal{G} \right)$ given that $\mathcal{C}$ and $\mathcal{G}$ together typically have of order $\sim 15$ free parameters \citep{Feroz_10}. Most importantly, due to the dependence on $\mathcal{C}$, this would require of order $\mathcal{O}(10^6)$ cosmological simulations which is computationally infeasible.
	
	\subsection{Surrogate model}
	
	A widely used approach is constructing a surrogate model for $\hat{\mathbf{D}} \left( \mathcal{C}, \mathcal{G} \right)$ that is computationally cheap to evaluate. First, $\mathcal{O} (10^2)$ cosmological simulations are run. For each simulaton, $\mathcal{O} (10^3)$ realisations of $\mathcal{G}$ are built and the predictions $\hat{\mathbf{D}} \left( \mathcal{C}, \mathcal{G} \right)$ computed. The values for $\mathcal{C}$ and $\mathcal{G}$ for those $\mathcal{O} (10^5)$ realisations of $\hat{\mathbf{D}} \left( \mathcal{C}, \mathcal{G} \right)$ are chosen to efficiently sample the prior space. Finally, the predictions $\hat{\mathbf{D}}$ are used as training points for an emulator, most commonly a Gaussian process emulator, to predict $\hat{\mathbf{D}}$ for arbitrary points in the cosmology and galaxy-halo parameter space without the need to re-run expensive simulations.
	
	The most extensive example of such an emulator approach in the context of the small-scale clustering of galaxies is the work of \cite{Zhai_19} \citep[also see][]{Kwan_15, Nishimichi_18, Wibking_19}. In this study, the authors constructed an emulator for the redshift-space clustering of galaxies in the BOSS CMASS survey. Using this work as an example, we will point out some of the shortcoming of the emulator approach. First, it is important to realise that a surrogate model such as Gaussian process emulation cannot perfectly reproduce the full forward-modelling approach. In the case of \cite{Zhai_19}, it was shown that emulator inaccuracies are roughly of the same order of magnitude as the typical observational uncertainties of the data. This has two important implications. First, these emulator inaccuracies will degrade or possibly bias our posterior inference. In the work of \cite{Zhai_19}, typical cosmological constraints are degraded by a factor of $\sim 1.5$ due to emulator noise. Secondly, the emulator accuracy will likely degrade further with increased dimensionality of $\mathcal{C} \bigotimes \mathcal{G}$. In principle, this could be mitigated by increasing the number of training points for the emulator. But, for example, the processing time of a Gaussian process emulator scales with the number of training points to the third power. This puts a practical limit on the number of training points and, thereby, the complexity of the input models for $\mathcal{C}$ and $\mathcal{G}$. For example, with the exception of \cite{Wibking_19} who use a simplistic linear Taylor expansion emulator, we are not aware of any emulator for the non-linear galaxy clustering that takes into account cosmology and galaxy assembly bias \citep[see e.g.][]{Zentner_14, Wechsler_18}. Several studies have shown that neglecting this effect in the modelling of $\mathcal{G}$ can lead to biased cosmological inferences \citep[see e.g.][]{Zentner_14, Wibking_19}. We note that \cite{Yuan_19b} have recently demonstrated that an emulator for galaxy assembly bias at fixed cosmology can be built.
	
	Finally, \cite{Wibking_20} have developed an emulator for the galaxy clustering and galaxy-galaxy lensing of BOSS galaxies. A novel feature of their approach is that instead of emulating the observables directly the authors emulate corrections to observables calculated from analytic halo models. It was shown that this method can increase emulator accuracy and thereby should allow more complex models of the galaxy-halo connection. Ultimately, this shows that it might be possible to develop the surrogate modelling approach further to overcome some of its shortcomings.

	\subsection{Cosmological evidence}
	
	Our goal here is to develop a new approach different from the emulation of observables. This new method easily allows for arbitrary models of the galaxy-halo connection and also performs the marginalisation over $\mathcal{G}$ in an exact manner, thereby increasing the accuracy of the posterior inference. Broadly speaking, our new approach utilises the fact that realisations of the galaxy-halo connection $\mathcal{G}$ on top of a cosmological simulation are computationally orders of magnitude cheaper than running the cosmological simulation itself. We start by integrating Eq.~\eqref{eq:bayes_full} over $\mathcal{G}$,
	\begin{equation}
	    \begin{split}
	        P\left( \mathcal{C} | \mathbf{D} \right) &= \int P\left( \mathcal{C}, \mathcal{G} | \mathbf{D} \right) \rmd \mathcal{G}\\
	        &= \frac{1}{\mathcal{Z}(\mathbf{D})} \int P\left( \mathbf{D} | \hat{\mathbf{D}} \left( \mathcal{C}, \mathcal{G} \right) \right) P(\mathcal{C}) P(\mathcal{G}) \rmd \mathcal{G}\\
	        &= \frac{\mathcal{Z} (\mathbf{D} | \mathcal{C}) P(\mathcal{C})}{\mathcal{Z}(\mathbf{D})} \, .
	        \label{eq:bayes_cosmology}
	   \end{split}
	\end{equation}
	In the above equation we have identified the Bayesian evidence for the observations $\mathbf{D}$ given a certain cosmological model $\mathcal{C}$,
	\begin{equation}
	    \mathcal{Z} (\mathbf{D} | \mathcal{C}) \equiv \int P\left( \mathbf{D} | \hat{\mathbf{D}} \left( \mathcal{C}, \mathcal{G} \right) \right) P(\mathcal{G}) \rmd \mathcal{G} \, .
	    \label{eq:evidence_cosmology}
	\end{equation}
	This shows that if we were able to calculate the Bayesian evidence for a given cosmological model, we would only need to interpolate $\mathcal{Z} (\mathbf{D} | \mathcal{C})$ instead of $\hat{\mathbf{D}} (\mathcal{C}, \mathcal{G})$. Let us denote the number of dimensions of $\hat{\mathbf{D}}$, $\mathcal{C}$ and $\mathcal{G}$ by $n_{\rm D}$, $n_{\rm C}$ and $n_{\rm G}$, respectively. Then, $\mathcal{Z} (\mathbf{D} | \mathcal{C})$ is a $\mathbb{R}^{n_{\rm C}} \rightarrow \mathbb{R}$ and $\hat{\mathbf{D}} (\mathcal{C}, \mathcal{G})$ a $\mathbb{R}^{n_{\rm C} + n_{\rm G}} \rightarrow \mathbb{R}^{n_{\rm D}}$ function. Typical numbers of $(n_{\rm D}, n_{\rm C}, n_{\rm G})$ are $(42, 5, 12)$ \citep{More_15} or $(37, 8, 7)$ \citep{Zhai_19}. In principle, $n_{\rm D}$ can be reduced in specific cases through principal component analysis \citep{Nishimichi_18} or by emulating fitting functions \citep{McClintock_19}. But even in such cases, using the cosmological evidence still results in a dramatic reduction in dimensionality. Moreover, the Bayesian evidence for a given cosmology has no explicit dependence any more on the parameters of the galaxy-halo connection. Thus, as long as we can efficiently compute the integral in Eq.~\eqref{eq:evidence_cosmology}, we can allow for arbitrarily complex models of the galaxy-halo connection. 
		
	Fortunately, there are efficient numerical methods like, for example, nested sampling \citep{Skilling_04} to calculate the complex multi-dimensional integral in Eq.~\eqref{eq:evidence_cosmology}. Generally, there are many possible ways to calculate $\hat{\mathbf{D}} (\mathcal{C}, \mathcal{G})$. The most direct method is to first populate a dark matter halo catalogue with galaxies according to the model $\mathcal{G}$. Afterwards, the summary statistic $\hat{\mathbf{D}} (\mathcal{C}, \mathcal{G})$ is measured directly from the mock galaxy catalogue \citep[][]{Zentner_19, Vakili_19, Zhai_19}. However, in this work, we make use of the method described in \cite{Zheng_16}. In this method, correlation functions between halos of different properties are computed. The expected correlation functions between galaxies, which will serve as our summary statistic, can then be obtained by convolving the halo abundances and correlation functions with the probabilities to host galaxies. This allows us to make very fast and accurate predictions for galaxy correlation functions compared to the traditional naive direct population method. In Section \ref{sec:application}, we demonstrate that it is computationally feasible to calculate the Bayesian evidence for a given cosmology and make use of the dramatic reduction in dimensionality achieved with CEM.

	\section{Model ingredients}
	\label{sec:ingredients}
	
	In this section, we outline several of the ingredients going into Eq.~\eqref{eq:evidence_cosmology}. Specifically, we outline how we sample the cosmological parameter space $\mathcal{C}$, our model for the galaxy-halo connection $\mathcal{G}$ and our observational data vector $\mathbf{D}$. Finally, we describe how we calculate the Bayesian evidence $\mathcal{Z} (\mathbf{D} | \mathcal{C})$ in practice.

	\subsection{Cosmology \texorpdfstring{$\mathcal{C}$}{C}}

	In this work, we use simulations from the Aemulus project \citep{DeRose_19} to sample the cosmological parameter space $\mathcal{C}$. This simulation suite consists of $75$ individual simulations which vary in $7$ cosmological parameters: the total matter density $\Omega_\rmm$, the baryon density $\Omega_\rmb$, the Hubble constant $H_0$, the spectral index $n_s$, the amplitude of matter fluctuations $\sigma_8$, the equation of state of dark energy $w_0$ and the effective number of neutrino species $N_{\rm eff}$. The combinations of these $7$ parameters are chosen to efficiently sample the $4 \sigma$ CMB \citep{Hinshaw_13, Planck_14} plus baryon acoustic oscillation constraints of \citet{Anderson_14}. Among the $75$ simulations, there are $7$ groups of $5$ simulations each that have identical cosmological parameters but different random seeds. We follow the nomenclature of \citet{DeRose_19} and call them ``test simulations'' since they have been used to asses the emulator accuracy.
	
	Each simulation uses $1400^3$ particles to trace structure formation in a cubic volume of $(1050 \Mpch)^3$, resulting in a particle mass of $m_\rmp = 3.51 \times 10^{10} (\Omega_\rmm / 0.3) \msunh$. Dark matter haloes are identified with the {\sc ROCKSTAR} halo finder \citep{Behroozi_13a}. From the publicly available halo catalogues, we use the $z=0.55$ snapshots and keep all field halos with a halo mass of $M_{200 \rmm} > 100 m_\rmp$ where $M_{200 \rmm}$ is the spherical overdensity mass with respect to $200$ times the matter background density. Such a selection is appropriate for studying galaxies in the BOSS CMASS galaxy sample \citep[see e.g.][]{Guo_15a}. We assign a concentration parameter $c_{200 \rmm} = r_{200 \rmm} / r_s$ using the relation between $c_{200 \rmm}$ and $V_{\rm max}$ for a spherical NFW halo \citep{Prada_12, Klypin_16}. The halo concentration is later used to assign phase-space coordinates to satellite galaxies and to probe galaxy assembly bias.

	\subsection{Galaxy-halo connection \texorpdfstring{$\mathcal{G}$}{G}}
	\label{subsec:galaxy-halo_connection}
	
	Dark matter haloes in the Aemulus simulation are populated using a Halo Occupation Distribution (HOD) approach. We assume that every halo can host at most one central galaxy and an unlimited number of satellite galaxies. More specifically, we use the HOD parameterisaton of \cite{Zheng_07}, combined with the decorated HOD (dHOD) generalisation introduced by \cite{Hearin_16}. In this model, the mean number of central galaxies hosted by a halo of mass $M$ is
	\begin{equation}
	    \langle N_{\rm cen} | M \rangle = \frac{1}{2} \left( 1 + \mathrm{erf} \left[ \frac{\log M - \log M_{\rm min}}{\sigma_{\log M}} \right] \right)
	\end{equation}
	with $\log M_{\rm min}$ and $\sigma_{\log M}$ as free parameters. On the other hand, the average number of satellites is given by
	\begin{equation}
	    \langle N_{\rm sat} | M \rangle = \left( \frac{M - M_0}{M_1} \right)^\alpha
	\end{equation}
    for $M > M_0$ and $0$ otherwise. Here, $M_0$, $M_1$ and $\alpha$ are free parameters.
    
    The dHOD framework of \cite{Hearin_16} allows to further modulate these average occupation numbers with a secondary halo parameter besides halo mass. The motivation behind this is that it is well-known that the clustering properties of dark matter haloes depend on other halo variables besides halo mass, an effect called halo assembly bias \citep[see e.g.][]{Gao_05, Wechsler_06}. If the occupation of dark matter haloes with galaxies depends on any of these secondary halo parameters, an effect called galaxy assembly bias, the clustering of galaxies will be affected. We choose the halo concentration $c_{200 \rmm}$ as the secondary halo variable because it strongly affects halo clustering \citep[e.g.,][]{Wechsler_06, Jing_07, Gao_07, Han_19} and because concentration strongly correlates with halo age \citep[e.g.,][]{Navarro_97, Wechsler_02, Ludlow_13} which, in turn, might correlate with galaxy properties \citep[e.g.,][]{Hearin_13c}. In this case, the average number of galaxies is given by
    \begin{equation}
        \langle N_{\rm gal} | M, c \rangle = \langle N_{\rm gal} | M \rangle \pm \delta N_{\rm gal},
        \label{eq:N_cen}
    \end{equation}
    where $\delta N_{\rm gal}$ is added if the concentration of a halo is larger than the median of all haloes at that mass and subtracted otherwise. For centrals, $\delta N_{\rm gal}$ is equal to
    \begin{equation}
        \delta N_{\rm cen} = A_{\rm cen} \left( 0.5 - \left| 0.5 - \langle N_{\rm cen} | M \rangle \right| \right),
    \end{equation}
    which fulfills the constraint $0 \leq \langle N_{\rm cen} | M, c \rangle \leq 1$ for $-1 \leq A_{\rm cen} \leq 1$ as a free parameter. On the other hand, for satellites we have
    \begin{equation}
        \delta N_{\rm sat} = A_{\rm sat} \langle N_{\rm sat} | M \rangle,
    \end{equation}
    where $-1 \leq A_{\rm sat} \leq 1$ is another free parameter. We assume that the numbers of centrals and satellites in individual haloes obey Bernoulli\footnote{A Bernoulli distribution is a discrete probability distribution with $0$ and $1$ as the only possible outcomes.} and Poisson distributions, respectively.
    
    Finally, we need to assign phase-space coordinates to central and satellite galaxies. Centrals are assigned the position and velocity of their host halo core. Specifically, since we use {\sc ROCKSTAR}, the position corresponds to that of the phase-space density peak and the velocity is equal to the average velocity of all dark matter particles within $0.1 r_{200 \rmm}$ around the phase-space density peak. On the other hand, satellite phase-space positions are drawn from a spherically symmetric NFW profile. We allow the radial distribution of satellites to have a different concentration parameter than the dark matter halo, $c_{\rm sat} = \eta \, c_{\rm dm}$, and treat $\eta$ as a free parameter in our modelling. The velocities of satellites are drawn from a local Gaussian velocity distribution without orbital anisotropy. The mean of the distribution is that of the host halo core and the second moment $\sigma$ is obtained by solving the spherically symmetric, time-independent Jeans equation \citep{vdBosch_04, More_11, Lange_19b}. Thus, for a satellite at distance $r$ in a halo with mass $M$, concentration $c$ and halo radius $r_\rmh$:
    \begin{equation}
		\sigma^2 = \frac{G \eta^2 c^2 M}{r_\rmh g(c)} \left( \frac{r}{r_\rmh} \right) \left( 1 + \frac{\eta c r}{r_\rmh} \right)^2 \int\limits_{\eta c r / r_\rmh}^{\infty} \frac{g(y / \eta)\mathrm{d}y}{y^3 (1 + y)^2},
		\label{eq:Jeans}
	\end{equation}
	with $g(x) = \ln (1 + x) - x / (1 + x)$. Altogether, as shown in Table~\ref{tab:priors}, we have $8$ free parameters to describe the galaxy-halo connection $\mathcal{G}$.
    	
	\subsection{Summary Statistic \texorpdfstring{$\mathbf{D}$}{D}}
	
	For the purpose of this paper, we take the number density of galaxies and the redshift-space multipole moments $\xi_\ell (s)$ as the observational data vector $\mathbf{D}$. The multipole moments depend on the correlation function $\xi (s, \mu)$ that measures the excess probability of pairs separated by $(s, \mu)$ over a random distribution. Here, $s = \sqrt{r_\rmp^2 + \pi^2}$ and $\mu = r_\rmp / s$, and $r_\rmp$ and $\pi$ are the projected and line-of-sight separations of galaxy pairs, respectively. The multipole moments of order $\ell$ are then defined via
	\begin{equation}
	    \xi_\ell (s) = \frac{2\ell + 1}{2} \int\limits_{-1}^{1} L_\ell (\mu) \xi (s, \mu) \rmd \mu \, ,
	\end{equation}
	\citep{Hamilton_92}. In the above equation $L_\ell$ represents the Legendre polynomial of order $\ell$. Throughout this work, we use the monopole $\xi_0$ and quadrupole moments $\xi_2$ as our observables. We have experimented with including the hexadecapole moments $\xi_4$, but find that they do not provide much additional constraining power.
	
	We measure the redshift-space correlation function for mock galaxy populations using {\sc halotools}, specifically the {\sc s\_mu\_tpcf} and {\sc tpcf\_multipole} functions. We use linear bins for $\mu$ going from $-1$ to $+1$ with $\Delta \mu = 0.05$ and $14$ logarithmic bins in $s$ spanning the range $\log(s/\Mpch) \in [-1.0,1,8]$, the same bins as in \cite{Guo_15a}. These functions assume the distant observer approximation, treating one of the Cartesian coordinates of the simulation as the line-of-sight direction; to reduce the noise, we measure the multipole moments $3$ times by assuming the line-of-sight direction to be aligned with the $x$, $y$ and $z$-axis and averaging the result.
	
	In principle, we could measure the multipole moments in a given simulation by directly populating dark matter haloes in the simulation with galaxies. However, this is computationally very expensive and comes with realisation noise due to the random number and phase-space positions of galaxies. Instead, we use a tabulation method \citep{Neistein_12, Reid_14, Zheng_16} to speed up the computation dramatically and eliminate any realisation noise. We first take all haloes in a given simulation to serve as tracers of central galaxies. We furthermore assign to each halo of mass $M$ a Poisson number of satellite tracers with expectation value $3 \times (M/10^{13} \Msunh)$. This expectation value is chosen to be significantly larger than the number of satellites we typically expect in haloes of that mass \citep[see e.g.][]{Guo_15a}. We then bin all haloes and their central and satellite tracers by halo mass and whether the concentration is above or below the median. Next, we measure all cross- and auto-correlation multipole moments between all tracers in each bin. One can then show that an estimate for the galaxy number density and the multipole moments of any arbitrary galaxy-halo model are given by
	\begin{equation}
	    \hat{n}_{\rm gal} = \sum\limits_{i = \rmc, \rms} \sum\limits_{k = 1}^{n_{\rm bins}} N_{\rmh, k} \langle N_i | M_k, c_k \rangle
	\end{equation}
	and
	\begin{equation}
	    \begin{split}
	        \hat{\xi}_\ell = \hat{n}_{\rm gal}^{-2} \sum\limits_{i = \rmc, \rms} \sum\limits_{j = \rmc, \rms} \sum\limits_{k = 1}^{n_{\rm bins}} \sum\limits_{l = 1}^{n_{\rm bins}} \Big[ & N_{\rmh, k} N_{\rmh, l} \langle N_i | M_k, c_k \rangle \\
	        & \langle N_j | M_l, c_l \rangle \xi_{\ell, kl}^{ij} \Big] \,,
	    \end{split}
	\end{equation}
	respectively. In the above expression, $N_{\rmh, k}$ denotes the number of haloes in bin $k$ and, for example, $\xi_{\ell, kl}^{\rmc \rms}$ denotes the multipole moments between centrals in bin $k$ and satellites in bin $l$. The above estimate $\hat{\xi}_\ell$ approaches the expectation value of $\xi_\ell$ for sufficiently small halo mass bins. We find that $100$ logarithmic bins in halo mass is sufficient to adequately sample all haloes with mass $M > 3.52 \times 10^{13} (\Omega_\rmm/0.3) \msunh$ (corresponding to 100 particles). With such a bin width of $\sim 0.03 \ \mathrm{dex}$, any biases in $\xi$ are less than $5\%$ of the observational uncertainty for a BOSS CMASS-like sample \citep[see][]{Guo_15a}. The above method only works for a fixed value of the satellite radial profile parameter $\eta$. In practice, it suffices to tabulate correlation function for bins in $\eta$ of $\Delta \log \eta = 0.1$ and linearly interpolate between them.
	
	Finally, when measuring correlation functions from galaxy surveys, we measure angular separations, $\theta$, and redshift differences, $\Delta z$, instead of $r_\rmp$ and $\pi$. Generally, measurements of correlation functions are made by assuming a reference cosmology $\mathcal{C}_{\rm ref}$ to translate $(\theta, \Delta z)$ into comoving coordinates $(r_\rmp, \pi)$. Thus, we need to correct our predictions made using a simulation with parameters $\mathcal{C}_{\rm sim}$ for such a reference cosmology. This is known as the Alcock-Paczynski (AP) effect and can be modelled through a simple re-scaling of the coordinates in the box:
    \begin{equation}
        r_{\rmp, \rm ref} = r_{\rmp, \rm sim} \frac{d_{\rm com} (z | \mathcal{C}_{\rm ref})}{d_{\rm com} (z | \mathcal{C}_{\rm sim})}
        \label{paper_5:eq:rp_stretch}
    \end{equation}
    and
    \begin{equation}
        \pi_{\rm ref} = \pi_{\rm sim} \frac{E (z | \mathcal{C}_{\rm sim})}{E (z | \mathcal{C}_{\rm ref})} \, .
        \label{paper_5:eq:pi_stretch}
    \end{equation}
    In the above expressions, $d_{\rm com}$ is the comoving distance in $\Mpch$ and $E(z) = \sqrt{\Omega_{\rmm, 0} (1 + z)^3 + (1 - \Omega_{\rmm, 0}) (1 + z)^{3(1 + w_0)}}$. Throughout this work we use the best-fit Planck15 ``TT,TE,EE+lowP+lensing+ext'' $\Lambda$CDM cosmology \citep{Planck_16}, i.e. $\Omega_{\rmm, 0} = 0.307$ and $w_0 = -1$, as our reference cosmology $\mathcal{C}_{\rm ref}$.
	
	\subsection{Bayesian evidence \texorpdfstring{$\mathcal{Z}$}{Z}}
	
	\begin{table}
        \centering
        \begin{tabular}{cccc}
            \hline
            Parameter & Minimum & Maximum & Mock Data\\
            \hline\hline
            $\log M_{\rm min}$ & $12.5$ & $14.0$ & $13.36$\\
            $\sigma_{\log M}$ & $0.2$ & $1.0$ & $0.65$\\
            $\log M_0$ & $12.0$ & $15.0$ & $13.28$\\
            $\log M_1$ & $13.5$ & $15.0$ & $14.21$\\
            $\alpha$ & $0.5$ & $2.0$ & $1.02$\\
            $A_{\rm cen}$ & $-1$ & $+1$ & $0$ or $+0.5$\\
            $A_{\rm sat}$ & $-0.4$ & $+0.4$ & $0$ or $-0.1$\\
            $\log \eta$ & $-0.5$ & $+0.3$ & $0$\\
            \hline
		\end{tabular}
		\caption{The prior ranges and input values for mock catalogues of all parameters describing the galaxy-halo connection $\mathcal{G}$.}
		\label{tab:priors}
	\end{table}
	
	Calculating the Bayesian evidence first requires a model of the likelihood $P(\mathbf{D} | \hat{\mathbf{D}} (\mathcal{C}, \mathcal{G}))$. In this work, we assume a multivariate Gaussian distribution for the correlation function multipole moments and an uncorrelated error of $\sigma_{n_{\rm gal}} = 0.05 n_{\rm gal}$ for the number density of galaxies:
	\begin{equation}
	    P(\mathbf{D} | \hat{\mathbf{D}} (\mathcal{C}, \mathcal{G})) \propto \frac{(n_{\rm gal} - \hat{n}_{\rm gal})^2}{2 \sigma_{n_{\rm gal}}^2} + \frac{1}{2} (\xi - \hat{\xi})^{\rm T} \Sigma^{-1 } (\xi - \hat{\xi}) \, .
	    \label{eq:likelihood}
	\end{equation}
    In the above equation, $\xi$ contains the monopole and quadrupole measurements, i.e. $\xi = [\xi_0, \xi_2]$. We use the covariance matrix $\Sigma$ from the volume-limited BOSS CMASS sample of \cite{Guo_15a} which also accounts for correlations between the monopole and quadropole measurements. Additionally, to calculate the integral in Eq.~\eqref{eq:evidence_cosmology}, we need to specify the prior distribution $P(\mathcal{G})$. In this work, we use flat priors as listed in Table~\ref{tab:priors}. Note that we limit the prior of $A_{\rm sat}$ to $\pm 0.4$ instead of $\pm 1$. The reason is that the latter would be a rather unphysical model in which low or high-concentration halos have $0$ satellites.
	
	Finally, to compute the multidimensional integral in Eq.~\eqref{eq:evidence_cosmology} me make use of nested sampling \citep{Skilling_04} as implemented in {\sc MultiNest} \citep{Feroz_08, Feroz_09}. The main idea of nested sampling is to replace the multi-dimensional integral over $\mathcal{G}$ with a one-dimensional integral along iso-likelihood surfaces. The algorithm starts by randomly drawing a certain number of so-called live points from the entire prior space and calculates their likelihood. Afterwards, new points are drawn from the parameter space $\mathcal{G}$ and their likelihood values are calculated. If the likelihood of a new proposed point is larger than the lowest likelihood in the active set, $\mathcal{L}_{\rm min}$, the new point replaces the lowest-likelihood point in the active set. The lowest likelihood point previously in the active set is added to the so-called inactive set. One can show that the volume probed by the active set decreases exponentially with each newly drawn active point. With this information, at any given iteration, we can use all points from the active and the inactive set to calculate the total Bayesian evidence $\mathcal{Z}$ and an associated uncertainty $\Delta \ln \mathcal{Z}$. In order for nested sampling to be feasible, proposed points are not drawn from the entire prior space but from a smaller volume that encompasses all points with $\mathcal{L} > \mathcal{L}_{\rm min}$. While this volume can be estimated from all points in the active set, it is crucial for the volume from which points are drawn to fully contain all points with $\mathcal{L} > \mathcal{L}_{\rm min}$. Thus, the volume from which proposed points are drawn should be larger than the volume of the active set, thereby reducing the sampling efficiency of proposed points that are accepted. In this work, we use $1000$ live points and a stopping criterion of $\Delta \ln \mathcal{Z} = 0.01$. Using more live points would decrease the statistical uncertainty in the evidence estimate slightly but increase the overall runtime. We also use a the target sampling efficiency to $0.1$ with constant efficency mode turned off. Using a lower efficiency does not affect our results while only increasing the runtime. Finally, we also use the importance nested sampling \citep{Feroz_13} mode of {\sc MultiNest}. In practice, once the correlation functions have been tabulated each evaluation of the Bayesian evidence takes only of order $0.3$ CPU hours on a single run-of-the-mill processor from 2014.
	
	\section{Application to mock catalogue}
	\label{sec:application}
	
	In this section, we demonstrate the viability of CEM outlined in Section~\ref{sec:methods}. We do so by applying it to mock observations with a known input cosmology $\mathcal{C}$. If the method works correctly, we should be able to recover the input cosmology from our analysis. To produce mock data, we take the $5$ realisations of the third test cosmology of the Aemulus simulation suite, i.e. ``T02'' in Table 2 of \citet{DeRose_19}. The reason for this choice is simply that its cosmological parameters lie roughly in the middle of all simulations. We assume the best-fit halo occupation model of \citet{Guo_15a} (see Table~\ref{tab:priors}) and average the predicted summary statistic for the $5$ realisations, each time averaging over the three line-of-sight projections. Note that the mock observations are free of noise from random galaxy realisations, and that cosmic variance is reduced by using a cosmological volume of $5 \, h^{-3} \, \mathrm{Gpc}^3$ compared to $0.78 \, h^{-3} \, \mathrm{Gpc}^3$ in the actual BOSS data used by \cite{Guo_15a}. Since this implies that the theoretical uncertainty associated with the mock predictions is much smaller than the observational uncertainty we assume \citep{Guo_15a}, we should be able to recover the input cosmology to within much less than $1 \sigma$ of the posterior. Finally, for the moment, in both the mock observations and the models, we neglect galaxy assembly bias, i.e. $A_{\rm cen} = A_{\rm sat} = 0$. In what follows, the $5$ simulations used to predict the mock observations are not used to infer cosmological parameters.
    
    \begin{figure}
		\centering
		\includegraphics[width=\columnwidth]{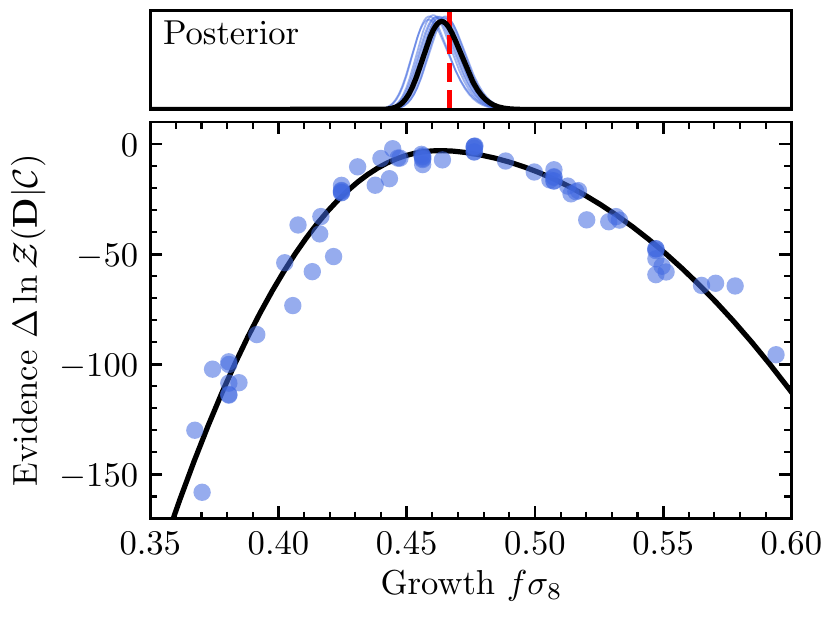}
		\caption{The statistical inferences on $f \sigma_8$ when analysing a mock catalogue. The lower panel shows the Bayesian evidence $\mathcal{Z} (\mathbf{D} | \mathcal{C})$ as a function of $f \sigma_8$. Each point corresponds to a simulation of the Aemulus simulation suite. Note that the normalisation of the Bayesian evidence is arbitrary and only the difference between cosmologies is relevant. We have thus chosen the highest evidence to correspond to $\ln \mathcal{Z} (\mathbf{D} | \mathcal{C}) = 0$. The solid lines shows our best-fit skew normal distribution to $\mathcal{Z} (\mathbf{D} | \mathcal{C})$. The smaller upper panel shows our inferred posterior constraint on $f \sigma_8$ and the dashed red line the ``true'' input value in the mock catalogue. Finally, the thin blue lines show random individual skew normal distributions from the posterior of fitting $\mathcal{Z} (\mathbf{D} | \mathcal{C})$.}
		\label{fig:fs8_vs_z}
	\end{figure}
    
    We start by calculating the Bayesian evidence $\mathcal{Z} (\mathbf{D} | \mathcal{C})$ for all the remaining $70$ cosmological simulations. In principle, the evidence is still a function of the $7$ cosmological parameters and it is difficult to gain an insight into a $7$-dimensional function from only $70$ simulations for $46$ independent cosmologies. Additionally, each value of $\mathcal{Z} (\mathbf{D} | \mathcal{C})$ suffers from cosmic variance, as discussed below. However, we expect redshift space multipoles to most strongly constrain $f \sigma_8$ where $f$ is the linear growth rate and both quantities are evaluated at the redshift of the simulation snapshots we use, $z = 0.55$. Fig.~\ref{fig:fs8_vs_z}, showing the Bayesian evidence $\mathcal{Z}$ as a function of $f \sigma_8$, indeed, suggests that the Bayesian evidence is primarily a function of $f \sigma_8$, i.e., almost all the variance in $\mathcal{Z}$ can be explained by its dependence on $f \sigma_8$.
	
	We now proceed to derive a posterior inference on cosmology. There are two problems we need to overcome. First, we only sample the cosmological parameter space with a very limited number of points. Secondly, every evidence estimate $\mathcal{Z}(\mathbf{D} | \mathcal{C})$ comes with an inherent uncertainty due to cosmic variance. We begin by estimating the uncertainty in our evidence estimate. Upon substituting  Eq.~\eqref{eq:likelihood} in Eq.~\eqref{eq:evidence_cosmology} it is clear that $\ln \mathcal{Z} \sim \chi_{\rm min}^2 / 2$, where $\chi_{\rm min}^2$ is the best-fit $\chi^2$. Let us assume that the prediction $\mathbf{D}$ from a simulation obeys a multivariate Gaussian with covariance $r_{\rm sim} \Sigma$. Here, $\Sigma$ is the covariance matrix of the observations and $r_{\rm sim}$ is a scaling factor that is related to the volume of the simulation with respect to the cosmological volume of the observations. Under these assumptions one can show that
	\begin{equation}
	    \Delta \ln \mathcal{Z} \approx \sqrt{\frac{r_{\rm sim}^2 N_{\rm D}}{2} + r_{\rm sim} \chi_{\rm min}^2}
	\end{equation}
	where $N_{\rm D}$ is the number of measurement points, i.e. $N_{\rm D} = 29$ in our case, and $\chi_{\rm min}^2$ is the best-fit $\chi^2$ at a given cosmology. We naively expect $r_{\rm sim} \approx V_{\rm obs} / V_{\rm sim}$ where $V_{\rm sim}$ and $V_{\rm obs}$ are the cosmological volumes of the simulations and the observations, respectively. However, $r_{\rm sim}$ is likely smaller because we project in all $3$ dimensions in redshift space and because we have a much better sampling of satellite galaxies. In practice, we leave $r_{\rm sim}$ as a free parameter to be determined empirically.
	
	In what follows, we will use an approach which broadly falls into the area of variational inference. That means, we approximate the non-trivial $\mathcal{Z}(\mathbf{D} | \mathcal{C})$ posterior distribution with an approximate analytic one. In particular, we assume that the evidence only depends on $f\sigma_8$ and that it follows a skew normal distribution,
	\begin{equation}
	    \begin{split}
	        \mathcal{Z} (\mathbf{D} | \mathcal{C}) &\approx \mathcal{Z} (\mathbf{D} | f \sigma_8) \\&\propto \left[ 1 + {\rm erf} \left( \frac{\alpha (f \sigma_8 - \mu)}{\sqrt{2} \sigma} \right) \right] \exp \left[ - \frac{(f \sigma_8 - \mu)^2}{2 \sigma^2} \right] \, .
	    \end{split}
	    \label{eq:evidence_fsigma8}
	\end{equation}
	The choice of this particular functional form for the cosmological evidence is motivated by the distribution of $\mathcal{Z}(\mathbf{D} | \mathcal{C})$ vs. $f \sigma_8$ shown in Fig.~\ref{fig:fs8_vs_z}. We note that we do not fit for the normalisation of $\mathcal{Z} (\mathbf{D} | f \sigma_8)$ or the total Bayesian evidence $\mathcal{Z} (\mathbf{D})$. Neither is necessary to derive the posterior $P(f \sigma_8 | D)$, i.e. Eq.~\eqref{eq:bayes_cosmology}, due to its normalisation condition, $\int P(f \sigma_8 | D) \mathrm{d} f \sigma_8 = 1$. Instead of just finding the best-fit skew normal distribution, we also consider the uncertainty from fitting the $\mathcal{Z}(\mathbf{D} | f \sigma_8)$ points. We choose flat priors for $\mu$, $\sigma$, $2 / \pi \arctan \alpha$ and $r_{\rm sim}$ with ranges $[0.35, 0.60]$, $[0.0, 0.1]$, $[-1, 1]$ and $[0, 1]$, respectively. Here, $\mu$, $\sigma$ and $\alpha$ are the parameters describing the location, scale and skewness of the skew normal distribution. Upon fitting $\mathcal{Z}(\mathbf{D} | f \sigma_8)$, we obtain a posterior sample of skew normal distributions $P(f \sigma_8 | D)$ that could fit the calculated evidence. We choose as the final cosmological posterior, the linear sum of all skew normal distributions in our posterior sample. In this way, our final cosmological inference accounts for uncertainties in fitting a functional form to the $\mathcal{Z} (\mathbf{D} | f \sigma_8)$ values.
	
	The upper panel in Fig.~\ref{fig:fs8_vs_z} shows inferred posterior constraint on $f \sigma_8$ when analysing the mock catalogue. We note that the $1\sigma$ range encompasses the input cosmology (indicated by the dashed, vertical line) despite this cosmology not being used in the analysis. This shows that the analysis was successful in recovering unbiased cosmological parameter  constraints. We have repeated this exercise with mock observations from the ``T01'' and ``T03'' simulations of the Aemulus simulation suite and obtained similar results. Overall, we can constrain $f \sigma_8$ to within $\sim 2\%$. We also note that the skew normal distribution fitted to the cosmological evidence is well constrained. For example, the mean of the skew normal is constrained to within $0.5\%$. In other words, the uncertainty in the mean of the posterior of $f \sigma_8$ due to the finite sampling of cosmology is of order $4$ times smaller than the posterior uncertainty in $f \sigma_8$ itself. Thus, our approach might be more precise than the emulation method which adds significant noise due to emulator errors \citep{Zhai_19}.
	
	We note that when fitting $\mathcal{Z} (\mathbf{D} | \mathcal{C})$ we neglected all cosmological parameters other than $f \sigma_8$. Generally, the cosmological evidence might very well depend on other parameters. However, the parameter space covered by the Aemulus simulation is too small to robustly uncover those. We find hints that $\mathcal{Z} (\mathbf{D} | \mathcal{C})$ might, for example, depend on parameters describing distortions due to the AP effect. However, the derived constraints are weak with respect to the Aemulus prior and taking this additional dependence into account has no appreciable impact on our $f \sigma_8$ constraint. Nevertheless, one should note that our constraints on $f \sigma_8$ are only strictly valid with the other cosmological parameters being in the Aemulus priors.
	
	\section{Impact of galaxy assembly bias}
	\label{sec:gab}
    
    \begin{figure}
		\centering
		\includegraphics[width=\columnwidth]{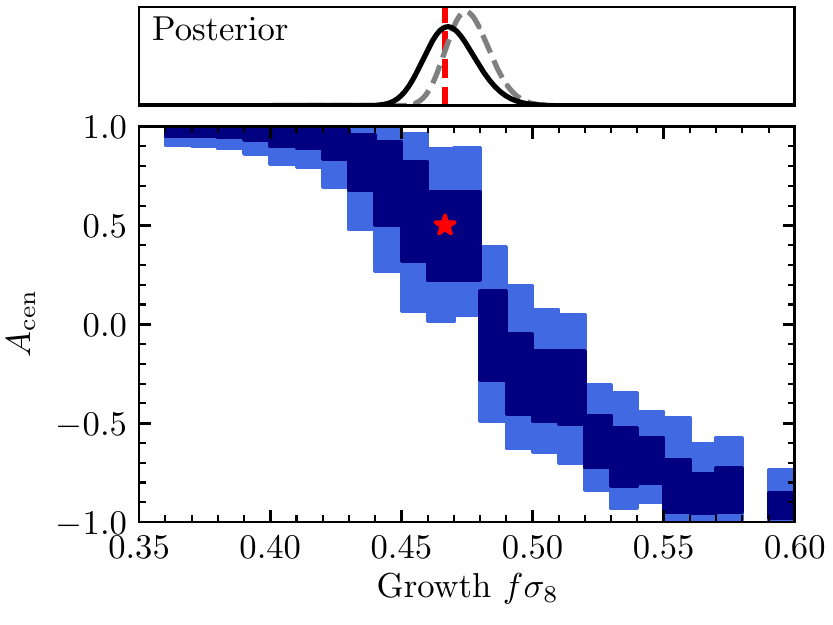}
		\caption{The impact of galaxy assembly bias on cosmological constraints. The upper panel shows posterior constraints on $f \sigma_8$. The solid (dashed) line represent results if galaxy assembly bias is included (neglected) in the analysis. In the lower panel, we show the $68\%$ and $95\%$ posterior constraints on $A_{\rm cen}$ in an analysis accounting for galaxy assembly bias. Each error bar corresponds to the average of all Aemulus simulations in that $f \sigma_8$ bin. In both panels, red elements denote the input values of the mock catalogue.}
		\label{fig:gab_vs_fs8_eps}
	\end{figure}
	
	So far, we have neglected galaxy assembly bias in the mock observations and the modelling. We now incorporate this effect into the mock observations by setting $A_{\rm cen} = +0.5$ and $A_{\rm sat} = -0.1$. This corresponds to a scenario in which high-concentration haloes have brighter centrals and fewer satellites. Such a scenario might be expected naturally because haloes with a higher concentration are older than their low-concentration counterparts. Thus, they had more time to tidally disrupt their satellites and thereby grow their centrals \citep[e.g.,][]{vdBosch_05b, Jiang_17}. All other HOD parameters are left unchanged. Next, we repeat the same analysis as in the previous section, this time also marginalising over both $A_{\rm cen}$ and $A_{\rm sat}$.
	
	The lower panel in Fig.~\ref{fig:gab_vs_fs8_eps} shows the posterior inferences about the assembly bias parameter $A_{\rm cen}$ that we obtain from analysing the mock data. Each error bar corresponds to one or more simulations of the Aemulus simulation suite, represented by their $f \sigma_8$ value. From this figure it is apparent that inferences about central galaxy assembly bias from small-scale redshift-space clustering depend strongly on the assumed cosmology. For example, for $f \sigma_8 \approx 0.43$ we could infer $A_{\rm cen} > +0.5$ at $99\%$ confidence, while adopting $f \sigma_8 \approx 0.49$ would result in inferring that $A_{\rm cen} < +0.5$ at $99\%$ confidence. Both results are clearly inconsistent with the true value of $A_{\rm cen} = +0.5$. Even more, both cosmological parameter combinations are still compatible with our cosmological inference at the $\sim 3 - 4\sigma$ level. Given that we have $29$ measurements and $8$ free parameters, we should expect to get a goodness of fit of $\chi^2 / {\rm dof} \approx 1.43 \pm 0.31$ and $1.76 \pm 0.31$ for $3\sigma$ and $4\sigma$ offsets, respectively. Thus, both values for $f \sigma_8$ might result in acceptable fits. This shows that even a good fit to observational data cannot reliably demonstrate the existence or absence of assembly bias unless one properly marginalises over the uncertainties in cosmology. Reassuringly, for the correct input cosmology, shown by the vertical red dashed line, we recover the correct galaxy assembly bias strength. On the other hand, we find the satellite galaxy assembly bias parameter $A_{\rm sat}$ to be virtually unconstrained unless $f \sigma_8$ is far from the input value.
	
	How does the presence of galaxy assembly bias affect cosmological constraints? Recall that we model the redshift-space clustering down to $0.1 \Mpch$. Recently, \cite{McCarthy_19} argued that galaxy assembly bias can only be ignored when modelling scales as large as $8 \Mpch$ and above. On these scales, the clustering seems to be determined solely by the large scale bias. Therefore, one does not need to model galaxy assembly bias due to its degeneracy with halo mass bias. However, \cite{McCarthy_19} show that the small-scale clustering cannot be predicted from the large-scale clustering unless one accounts for galaxy assembly bias. Thus, we do expect the presence of assembly bias to affect cosmological parameter constraints if it is not modelled. The solid lines in the upper panel of Fig.~\ref{fig:gab_vs_fs8_eps}, show the posterior inference on $f \sigma_8$ when allowing for galaxy assembly bias in calculating $\mathcal{Z} (\mathbf{D} | \mathcal{C})$. On the other hand, the dashed line shows the result if galaxy assembly bias is not accounted for. It is evident that not including galaxy assembly bias in the modelling leads to a small shift in the posterior by roughly $+0.01$ in $f \sigma_8$. This shift can be explained by the apparent degeneracy between $f \sigma_8$ and $A_{\rm cen}$: excluding assembly bias in the modelling corresponds to $A_{\rm cen} = 0,$ and in an analysis with galaxy assembly bias, $+0.01$ in $f \sigma_8$ is where $A_{\rm cen} \approx 0$ is preferred. We also note that the posterior uncertainty on $f \sigma_8$ does not increase substantially when including galaxy assembly bias in the analysis. Ultimately, this investigation shows that it is important to account for galaxy assembly bias in order to obtain unbiased inferences on cosmology. However, it seems more important to consider uncertainties in cosmological parameters when aiming for unbiased constraints on galaxy assembly bias, than vice versa.
	
	\section{Conclusion}
	\label{sec:conclusion}
	
	We have presented a new method to derive cosmological parameter constraints from the small-scale distribution of galaxies and matter. This method, which relies on numerical simulations, presents an alternative to using surrogate models for summary statistics as a function of cosmology and the galaxy-halo connection \citep[see e.g.][]{Heitmann_16, Nishimichi_18, DeRose_19, Zhai_19, Wibking_19}. The algorithm can be described as follows:
	\begin{enumerate}
	    \item Separate the full parameter space into the cosmological and the galaxy-halo connection parameter space. Realisations of the latter on top of a chosen cosmological parameter set, i.e. a cosmological simulation, are computationally inexpensive.
	    \item Perform the marginalisation over the galaxy-halo connection parameter space by calculating the Bayesian evidence $\mathcal{Z} (\mathbf{D} | \mathcal{C})$ for each cosmological simulation. Tabulate the values for the cosmological evidence and parameters of all simulations.
	    \item Determine a model for the cosmological evidence $\mathcal{Z} (\mathbf{D} | \mathcal{C})$ as a function of cosmological parameters.
	\end{enumerate}
	Compared to the emulation approach our CEM method based on cosmological evidence has several advantages.
	\begin{itemize}
	    \item CEM makes it easier to marginalise over highly complex models of the galaxy-halo connection, including those with galaxy assembly bias. On the other hand, for surrogate modelling, the complexity of the galaxy-halo connection is effectively limited by the accuracy of the emulator.
	    \item The marginalisation over the galaxy-halo connection, at least in the implementation presented here, is basically free of numerical noise, leading to more precise cosmological parameter inferences. This is different from the emulation method where even at a fixed cosmology errors stemming from emulating the dependence on the galaxy-halo connection could lead to inflated posteriors.
	    \item The determination of cosmology constraints can be further simplified by projecting the cosmological evidence into a suitable coordinate system. For example, in the case of the redshift-space clustering of BOSS CMASS galaxies and the Aemulus simulation suite, it is sufficient to model the evidence as a function of a single cosmological parameter. This additional reduction in the dimensionality cannot necessarily be applied to surrogate modelling.\footnote{A simple example would be to constrain cosmological parameters with just the number density of galaxies, $n_{\rm gal}$. Clearly, $\mathcal{Z} (n_{\rm gal} | \mathcal{C})$ should not depend on cosmology since $n_{\rm gal}$ does not contain cosmological information after marginalising over the galaxy-halo connection $\mathcal{G}$. On the other hand, the observable $\hat{n}_{\rm gal}$ at fixed values for $\mathcal{G}$ will depend on cosmology $\mathcal{C}$ and would have to be emulated as a function of both $\mathcal{G}$ and $\mathcal{C}$.}
	    \item CEM is a very transparent method: the main result of this approach, the cosmological evidence $\mathcal{Z} (\mathbf{D} | \mathcal{C})$ for different simulations with different cosmological parameters, can easily be included in a table accompanying an analysis. This would allow other groups to build their own models for how $\mathcal{Z}$ depends on cosmological parameters. One would not be stuck with the result from a particular set of assumptions and functional forms, like for example Eq.~\eqref{eq:evidence_fsigma8}. 
	    \item We demonstrated that CEM can be computationally cheap. For the analysis performed in this work, we only need roughly $200$ CPU hours to derive cosmological parameter constraints. The majority of that time is spent on tabulating the correlation functions. Deriving new cosmological constraints under a different model for the galaxy-halo connection takes only of order $20$ CPU hours on a standard server processor from 2014.
	\end{itemize}
	The exact performance of the CEM method also depends on how steps (ii) and (iii) are implemented. The implementations used in this paper are not the only possible choices. For example, it is important to acknowledge a shortcoming of our use of variational inference in step (iii): the assumed functional form for $\mathcal{Z} (\mathbf{D} | \mathcal{C})$ is only an approximation of the true cosmological evidence. It might be difficult to rigorously quantify systematic errors associated with the assumed functional form for $\mathcal{Z} (\mathbf{D} | \mathcal{C})$. This is especially important once the dependence on more cosmological parameters would need to be modelled. Finally, we want to point out that CEM and surrogate modelling should not necessarily be seen as mutually exclusive. For example, it might be possible to use surrogate modelling of observables at a fixed cosmology \citep[e.g.][]{Wang_15} in step (ii).
	
	We have tested the new CEM method on mock observations of the redshift-space clustering, specifically the monopole and quadrupole moments, of BOSS CMASS galaxies down to highly non-linear scales, $s = 0.1 \Mpch$. We have used simulations of the Aemulus simulation suite to forecast cosmological parameter constraints. We find that we can obtain a $\sim 2\%$ measurement of the cosmological growth rate $f \sigma_8$. We have also investigated how galaxy assembly bias interplays with cosmological parameters. We show that cosmological inferences can incur a small bias by not including galaxy assembly bias in the modelling. On the other hand, inferences about galaxy assembly bias from modelling clustering strongly depend on cosmology. This casts some doubt on inferences about the presence or absence of assembly bias from clustering studies at fixed cosmology \citep[see e.g.][]{Vakili_19, Zentner_19, Wang_19}.
	
	Our analysis here should also be compared to the results of \cite{Reid_14}. In this study, the authors model the redshift-space clustering down to $0.8 \Mpch$ and obtain a $2.5\%$ constraint on $f \sigma_8$. One of the main differences between \cite{Reid_14} and our analysis is that \cite{Reid_14} work with only a three simulations. To model the dependence on $f \sigma_8$, all halo velocities in the simulation are scaled by a factor $\gamma_{\rm HV}$ and interpreted as a corresponding change in $f \sigma_8$. However, \cite{Zhai_19} argue that such a method might underestimate the true uncertainty in $f \sigma_8$. Since we do not model $\gamma_{\rm HV}$, we cannot comment on this here. On the other hand, our analysis at least suggests that the fact that \cite{Reid_14} neglected galaxy assembly bias should not have strongly biased their conclusions.
	
	We expect to apply CEM to observational data in the near future. For example, one might constrain cosmological parameters from the redshift-space clustering \citep{Guo_15a} and galaxy-galaxy lensing \citep{Leauthaud_17, Lange_19c} of BOSS CMASS galaxies. However, we note that one first needs to verify that current models of the galaxy-halo connection are complex enough to accommodate realistic galaxy populations. This could be tested by running analysis on mock galaxy catalogues obtained from semi-analytic models or hydrodynamical simulations. For example, our analysis neglects parameters for central and satellite velocity bias commonly employed \citep[see e.g.][]{Tinker_06, Reid_14, Zhai_19}. We assume that centrals are located at the halo core which should greatly reduce the need for additional velocity offsets \citep{Behroozi_13a, Guo_16, Ye_17}. Similarly, satellite velocities are derived from solving the Jeans equation which allows physically motivated differences between the velocities of satellites and dark matter \citep{Lange_19a}. While these models are well motivated, it is still possible that including additional velocity bias parameters is required to obtain unbiased cosmological parameters. We believe that the method presented here will provide a valuable tool for these future works.
    
	\section*{Acknowledgements}

	We thank Jeremy Tinker and Zhongxu Zhai for helpful discussions regarding this work. We also thank the Aemulus collaboration for making their results public and giving us preliminary access.
	
	FvdB and JUL are supported by the US National Science Foundation (NSF) through grant AST 1516962. This research was supported  by the HPC facilities operated by, and the staff of, the Yale Center for Research Computing. FvdB received additional support from the Klaus Tschira foundation, and from the National Aeronautics and Space Administration through Grant No. 17-ATP17-0028 issued as part of the Astrophysics Theory Program. ARZ and KW are supported by the NSF through grant AST 1517563 and by the Pittsburgh Particle Physics Astrophysics and Cosmology Center (PITT PACC) at the University of Pittsburgh. HG is supported by the National Key Basic Research Program of China (No. 2015CB857003) and national science foundation of China (Nos. 11833005, 11828302, 11773049). Work done at Argonne National Laboratory was supported under the DOE contract DE-AC02-06CH11357.
	
	This work made use of the following software packages: {\sc matplotlib} \citep{Hunter_07}, {\sc SciPy} \citep{Jones_01}, {\sc NumPy} \citep{vdWalt_11}, {\sc Astropy} \citep{Astropy_13}, {\sc Corner} \citep{Foreman-Mackey_16}, {\sc MultiNest} \citep{Feroz_08,Feroz_09, Feroz_13}, {\sc PyMultiNest} \citep{Buchner_14}, {\sc Colossus} \citep{Diemer_18}, {\sc Spyder}, {\sc KBibTex} and {\sc Kile}.
	
	\bibliographystyle{mnras}
	\bibliography{bibliography}
	
	\label{lastpage}

\end{document}